\documentclass[11pt]{llncs}
\begin{document}

\title{Remarks on ``Toward Compact Interdomain Routing''}
\author{Victor S. Grishchenko}
\institute {Ural State University \\ \email{gritzko@ural.ru} }
\maketitle

\begin {abstract}
This paper critically examines some propositions and arguments of \cite{toward} regarding applicability of hierarchical routing and perspectives of compact routing. Arguments against the former are found to be inaccurate while the latter is found to be equivalent to well-known deployed solutions. Also, multiple (stacked) application of compact-routing solutions is found to be equivalent to hierarchical routing.
\end {abstract}

\section {Hierarchical routing}

By hierarchical routing (in general) I mean a family of algorithms, which 
\begin{itemize}
\item divide network into partitions, partitions into sub-partitions etc etc (top-bottom view) or vice-versa (bottom-up view) group routers into aggregates, aggregates into higher aggregates etc etc.
\item take routing decisions in terms of aggregates and thus, each node employs routing tables of logarithmical size.
\end{itemize}

Authors of \cite{toward} argument inapplicability of hierarchical routing by the fact that inter-AS connectivity topology has small-world features, i.e. ``it has essentially no remote nodes''. Thus, hierarchic network partitioning will be ineffective because it supposes that most of the nodes are ``remote''.

This argument is not valid on the following reason: hierarchical routing employs routing tables of logarithmical size, at any node (i.e. $\sim \log N$, where $N$ is the total number of nodes in the network). In fact, autonomous systems (ASes) aren't technically necessary for hierarchical routing to function because almost any node can afford routing tables of logarithmical size.
So, inapplicability of hierarchic routing schemes to ASes is a fault of ASes, not of hierarchic routing.

This problem has one more aspect. Autonomous systems are organizational entities without any forced geographic correlations. Geographically, AS may cover a country or a single device. 
Different from ASes, router-level internet topology is influenced by such geographic factors as distance and locality. Consider that RTT depends on distance or that 100km cable has much higher OPEX and CAPEX than 100m cable, etc etc. For any given IP device the overwhelming majority of all the IP devices are ``remote'' by any means: they are thousands of kilometers away, separated by many hops, etc etc. One may argue e.g. that router-level topology of the Internet is also scale-free \cite{powerlaw} so the diameter is logarithmic although it is bigger than AS graph diameter by some factor.  This approach has a problem of treating 1m hop and 100km hop as equals. If we will weight router connectivity graphs with round trip times then we will unsurprisingly find that space is still there \cite{distance,mine}.
So, locality is mostly lost when one takes AS-level hop-measured approach to the topology of the Internet.

So, arguments of \cite{toward} on inapplicability of hierarchical routing to the Internet are inaccurate.

\section {Compact routing}

\subsection {Compact routing is here and working}

By compact routing scheme I mean a solution which has
\begin{itemize}
\item sublinear per-node memory upper bound 
\item constant stretch upper bound 
\item node labels of logarithmic size
\end{itemize}

In particular, Thorup-Zwick \cite{tz} scheme of compact routing assumes that we split a network into $O(\sqrt{N})$ vicinities each of size $O(\sqrt{N})$. By routing a message first to the respective vicinity of the destination node and then to the destination node itself we may limit ourselves to routing tables of size $O(\sqrt{N})$. The side effect is the stretch of 3.

I will nickname this general approach as ``Dijkstra splitted into halves''. This compact routing scheme \cite{tz} ``splits'' a choice among $N$ variants into two orthogonal choices of $O(\sqrt{N})$ variants each (inter-vicinity and intra-vicinity routing). Still, both inter-vicinity and intra-vicinity routing remain to be Dijkstra shortest path routing.

This scheme is obviously equivalent to the current tandem of BGP and OSPF \cite{bgp,ospf}. Indeed, how many ASes exist today? An order of $10^{4}$ \cite{toward}, i.e. two bytes. IPv4 address is 4 bytes long. Thus, an order of complexity of intra-domain routing is the remaining 2 bytes. Both BGP and OSPF employ different flavours of Dijkstra's shortest path routing (with some exceptions, such as administrative policies or OSPF ``areas'', see the next section on ``split further'' schemes).

So, BGP+OSPF tandem splits a choice among $2^{32}$ variants into two orthogonal choices of $\sim 2^{16}$ variants each for the overhead of some stretch. It is effectively the desired compact routing.

\subsection{Very-very compact routing} \label{sec:very}

It is rather interesting, that stacking one compact routing solution on top of another we may ideally get a hierarchical routing scheme.

We may further apply ``compactization'' to both ``pieces of Dijkstra'', i.e. to aggregate vicinities into super-vicinities and to split each vicinity into sub-vicinities. This leads us to total per-node routing table size of $4N^{\frac{1}{4}}$, i.e. 4 tables per node $O(N^\frac{1}{4})$ each. It is rather obvious that applying this method again and again we may get size of each routing table under some pre-given threshold $k$. At the same time, the total number of routing tables increases. The resulting amount of routing information at an average node has an order of $2^i \cdot  N^{\frac{1}{2^i}} = log_{k}{N} \cdot N^{\log_{N}k} = log_{k}{N}~\cdot~k$, i.e. this approach guarantees logarithmic size of routing tables.

Each level of compactization replaces each ``shortest path step'' of a routed path with two ``shortest path little steps'': from a source to a respective landmark and then from the landmark to the destination. E.g. simple Dijkstra algorithm uses one such step, Thorup-Zwick uses two steps, Thorup-Zwick over AS graph makes about four steps: from the source to some landmark AS, then to the target AS, then to respective OSPF area, then to the destination (some straightforward case is assumed; in fact, intra--AS routing may proceed in a quite different way).

So, each compactization level adds a multiplicative stretch factor (3 for Thorup-Zwick). Thus, we may expect the total all-levels stretch to have an order of $3^i = (2^i)^{\log_23} = log_k^{\log_23}N$. For the case of IPv4 we have $N=2^{32}$, say $k=2^8$ \cite{cril}, so $i=2$ and the stretch incurred by double Thorup-Zwick compactization has an upper bound of $9$.



\subsection {Applying compact routing}

There are two basic scenarios of applying compact routing to the Internet: either \emph{instead} of ASes or \emph{on top} of ASes.
In the former case we'll get what we already have.
The latter case has no straightforward solution because there is no easy way to aggregate IPv4 prefixes into vicinities. Enlisting all prefixes leads us back to the problem of linear routing table growth. Another approach is probably some name-independent compact routing scheme which will allow an average BGP device to store $O(\sqrt{n}) = O(N^{\frac{1}{4}})$ entries, where $n$ is the number of announced prefixes (say, two bytes) and $N$ is the total number of devices or addresses (four bytes). I.e. the size of a routing table in the core of the network will be approximately equal to ``one byte'', an order of $2^8$ entries -- this result is in agreement with \cite{cril}.

This way we'll get exactly the case of stacked compact routing schemes (see Sec.~\ref{sec:very}). One may object that we don't harm much by introducing an additional \emph{average} stretch factor of e.g. $1.1$. The case is not that simple. An extra AS in the path may mean that an actual packet will travel over an ocean and back again. So, the average AS hop stretch factor of $1.1$ may contribute much more in physical parameters, such as RTT stretch. Of course, the reverse situation is also possible.

Besides the inevitable growth of stretch, that name-in\-de\-pen\-dent compact routing scheme deployed on top of ASes will be subject to some considerations of \cite{rfc3439}, namely the complexity-robustness spiral effect. One additional layer of routing infrastructure will increase overall system complexity.
It is highly probable, that users will try to introduce some hacks and workarounds to overcome cases of obviously suboptimal behavior (i.e. the inter-AS routing stretch mentioned). This will again add complexity and thus decrease reliability and robustness\ldots ~and so on and so forth.

\section{Conclusions}

\begin{description}
\item[First,] compact routing solutions are not anything new, because current routing infrastructure employs exactly this model of routing. 
\item[Second,] the impressive result of \cite{cril} (i.e. core routing table of $\sim 50$ entries) has more of hierarchical than of compact nature. More precisely, it is a stacked compact routing scheme.
\item[Third,] the kind of dependence between AS hop stretch and physical (say RTT) stretch is unclear. So, we can not precisely estimate effects of inter-domain stretch. 
\item[Fourth,] one more compactization layer will contribute to the complexity spiral effect by all means. 
\item[Fifth,] there is no straighforward way to implement name-dependent compact routing on top of ASes (in fact, to make interdomain routing ``more hierarchical'').
\end{description}


As far as I see it, one may take more fundamental approach to the problems of Internet routing by focusing on the following problems:
\begin{enumerate}
\item The metrics widely used in routing research oversee space and distance. IP hop or AS hop may easily vary by three to five orders of magnitude in terms of time, distance and cost. This logical flaw creates an illusion of a small world.
Our planet is rather small, but not to that extent.
\item Current IP addresses are not in sync with the actual network topology to the extent that sometimes researchers tend to interpret IPs as random identifiers (e.g. authors of \cite{toward} put much of their expectations into name-independent routing schemes).
\end{enumerate}
Solutions to these problems if found, will introduce a fresh look on the problems of the Internet routing infrastructure.

\section{Acknowledgments}

Author thanks everyone who tolerates his broken English.

\end{document}